\documentclass[aps, 10pt, prd, notitlepage,
twocolumn, superscriptaddress, nofootinbib, tightenlines]{revtex4-1}

\usepackage[linktocpage,breaklinks]{hyperref}
\usepackage[usenames,dvipsnames]{xcolor}
\usepackage{times}
\usepackage{newtxtext}
\usepackage{newtxmath}
\usepackage[T1]{fontenc}
\usepackage{amsmath}
\usepackage{tensor}
\usepackage[utf8]{inputenc}
\usepackage{mathrsfs}
\usepackage{bm}

\usepackage{graphicx}
\usepackage{epsfig}
\usepackage{epstopdf}

\usepackage{natbib}
\usepackage{hyperref}
\usepackage{cleveref}

\definecolor{romared}{RGB}{142,0,28}
\hypersetup{colorlinks=true,
            citecolor=romared,
            linkcolor=romared,
            urlcolor=romared}

\newcommand{\be}{\begin{equation}}
\newcommand{\ee}{\end{equation}}
\newcommand{\op}{\left}
\newcommand{\cl}{\right}

\begin{document}
\title{Dynamical signatures of black holes in massive Chern-Simons gravity: Quasibound modes and time evolution}

\begin{abstract}

Dynamical Chern-Simons (dCS) gravity is a promising extension of general relativity (GR), arising naturally from the low-energy limit of some string motivated theories. Even though dCS possesses an additional scalar degree of freedom, interestingly, the Schwarzschild black hole is an exact solution of it. 
Concerning dynamical phenomena, however, gravitational and scalar perturbations couple with each other, generating possible scenarios to understand the differences between dCS and GR. 
We study dynamical signatures of dCS considering that the scalar field potential has a mass term. 
We analyze the influence of the theory's parameter into the quasibound states and in the time evolution of purely gravitational initial profiles. 
We find that the coupling can make the dipolar modes \textit{less stable} and that at late times initial gravitational perturbations become contaminated with the scalar sector, presenting an oscillation that depends on the quasibound state frequencies. 
These results may be interesting in the light of superradiant instabilities in rotating systems and to find signatures of alternative theories in gravitational wave detections.

\end{abstract}

\author{\sc{Caio F. B. Macedo}}
\email{caiomacedo@ufpa.br}
\affiliation{Campus Salin\'opolis, Universidade Federal do Par\'a,
Salin\'opolis, Par\'a, 68721-000 Brazil}

%

\date{{\today}}
\maketitle

\section{Introduction}
\label{sec:int}

The gravitational wave detections by the LIGO and Virgo collaborations opened the door to test the strong field regime of gravity~\cite{Abbott:2016blz,Abbott:2016nmj,Abbott:2017oio,Abbott:2017vtc,TheLIGOScientific:2017qsa,Abbott:2017gyy}. In this regime, especially in the merger of compact objects, general relativity (GR) shows all its nuances and one has to study it through the use of full numerical relativity machinery. Additionally, with the possibility of GR not being the true theory of gravity, these scenarios are ideal to observe signatures of alternative theories of gravity~\cite{TheLIGOScientific:2016src} (see also~\cite{Konoplya:2016pmh,Yunes:2016jcc,Berti:2015itd,Cardoso:2016ryw,Berti:2018cxi,Berti:2018vdi,Barack:2018yly}). 

Since the discovery of GR, a plethora of alternative theories of gravity have been proposed, many of them motivated by problems related to the nonrenormalizabity of GR. An interesting class of alternative theories of gravity is the dynamical Chern-Simons (dCS) theory (see Ref.~\cite{Alexander:2009tp} for a review). The theory was introduced by Jackiw and Pi~\cite{Jackiw:1990mb}, but it can also arise from heterotic superstring scenarios~\cite{Campbell:1990fu,Campbell:1992hc}. We can write the dCS action as (we will use natural units $G=c=\hbar=1$)\footnote{We could also add a constant in the scalar field action, but this can be absorbed in a redefinition of  the scalar field~\cite{Pani:2011xj}.}
\begin{align}
S&=\frac{1}{16\pi}\int d^4x\sqrt{-g}\Big\{R
+\frac{\alpha}{4}\varphi\,{^*RR}\nonumber\\
&-\frac{1}{2}[g^{ab}\nabla_a\varphi\nabla_b\varphi+V_s(\varphi)]\Big\}+S_{\rm mat},
\end{align}
where $g$ is the metric determinant, $R$ is the Ricci scalar, $\varphi$ is the scalar field,
\be
{^*RR}=R_{abcd}^{*}R^{bacd}=\frac{1}{2}R_{abcd}\epsilon^{baef}{R^{cd}}_{ef},
\ee
the Pontryagin density, and $V_s(\varphi)$ is the scalar field potential. The dCS theory can also be thought of as a subclass of a more general class of theories of gravity called quadratic gravity~\cite{Alexander:2009tp,Yunes:2011we,Pani:2011gy}. Usually, the scalar field potential is set to zero, but here we will consider it to describe a massive scalar, i.e.,
\be
V_s=\mu^2|\varphi|^2,
\ee
where $\mu$ is the scalar field mass. The consideration of a massive scalar field can be thought of as a dominant term in an expansion of the potential $V_s$ in powers of $\varphi$. As such, we will consider it to be relatively small, scaled with the black hole (BH) mass as $M\mu\leq 0.5$. 
Moreover, small mass ranges are particularly interesting when rotation is considered, as the superradiant instability is highly suppressed for large $\mu M$~\cite{Zouros:1979iw,Dolan:2007mj}.

We will investigate dynamical features of massive dCS by using perturbation theory. Because the Pontryagin density vanishes for spherically symmetric spacetimes~\cite{Grumiller:2007rv}, spherical BHs are still described by the Scwarzschild spacetime, i.e.,
\be
ds^2=-f(r)dt^2+f(r)^{-1}dr^2+r^2d\Omega^2,
\ee 
where $f(r)=1-2M/r$.

Metric and scalar field perturbations of BHs in dCS gravity can be studied by using the standard formalism introduced by Regge-Wheller~\cite{Regge:1957td} and Zerilli~\cite{Zerilli:1970se}. Perturbations of spherically symmetric BHs in dCS were studied in the past~\cite{Alexander:2009tp,Molina:2010fb,Pani:2011xj}, and we will closely follow the definitions of Ref.~\cite{Pani:2011gy}. One can find that the even sector of the perturbation is precisely the same as that in GR, as scalar field perturbations couple only with the axial sector. Therefore, we will only study the axial sector in this paper.  Within this picture, one can compute the quasinormal modes (QNMs) of the system, which are natural frequencies of the BH, behaving as purely ingoing waves at the horizon and outgoing waves at infinity~\cite{Berti:2009kk}.

The QNMs in massless dCS was investigated in detail in Ref.~\cite{Molina:2010fb}, where the behavior of the mode frequencies were analyzed for a wide range of the parameter space. Additionally, the authors briefly discussed the influence of the mass term into the QNMs of static BHs in dCS. 
Essentially,  with the influence of the scalar mass, the behavior of the scalar modes as a function of the coupling constant is analogous to the massless case. Moreover, the gravitational mode dependence on the scalar mass is very mild. However, a study of one important part of the spectrum that arises when the mass term is present is still lacking: the quasibound (QB) states~\cite{Deruelle:1974zy,Damour:1976kh}. QB states are localized long-lived solutions, 
slowly leaking to the BH\footnote{We will see that in the dCS the 
QB mode will also leak to infinity mainly through the gravitational channel [see Eq.~\eqref{eq:binf2}].}. 
This part of the spectrum is important because it is responsible 
for the superradiant instabilities rotating spacetime solutions~\cite{Dolan:2007mj}.

In this work, we explore the QB states of massive dCS gravity, considering as the background the Schwarzschild spacetime. We compute the monopolar and dipolar modes, considering different values for the coupling constant. We show that the coupling impacts mostly the imaginary part of the modes, influencing, therefore, the decay time of initial perturbations that excites the modes. We obtain in this frequency window that the spacetime is stable, corroborating previous studies in the literature~\cite{Molina:2010fb,Kimura:2018nxk}. We also study how the mass term affects the time evolution of initial signals. We perform time evolutions of Gaussian wave packets, showing that the gravitational sector of the perturbations presents, at late times, a behavior similar to that of massive scalars.

The remainder of this paper is organized as follows. In Sec.~\ref{sec:pert}, we present the equations describing axial perturbations in massive dCS gravity. We present the direct integration method we will use to find the QB frequencies and the Breit-Wigner method to verify the results. We also present the setup for the time evolution of initial Gaussian profiles. In Sec.~\ref{sec:results}, we present the numerical results for the QB modes and the results for the time evolution of gravitational and scalar perturbations. Finally, in Sec.~\ref{sec:conclusion}, we present our discussion and conclusions.

\section{Perturbations of BHs in massive dCS gravity}
\label{sec:pert}

The equations describing axial perturbations of Schwarzschild black holes in massive dCS can be written as\footnote{See Ref.~\cite{Pani:2011gy} for a derivation.}
\be
\op(\frac{\partial^2}{\partial r_\star^2}-\frac{\partial^2}{\partial t^2}-{\bf V}\cl){\bf \Psi}(t,r_\star)=0,
\label{eq:perturbation}
\ee
where $r_\star$ is the tortoise coordinates, defined by $dr_\star=f(r)^{-1}dr$, and ${\bf \Psi}=\{\psi_g,\Theta\}$, in which $\psi_g$ represents the gravitational part of the perturbations and $\Theta$ is the scalar one. The potential $\bf V$ is given by
\begin{equation}
{\bf V}=\op[
\begin{array}{cc}
	V_{11} & V_{12}\\
	V_{21} & V_{22}
\end{array}\cl],
\label{eq:system}
\end{equation}
with
\begin{align}
	V_{11}&=f\left(\frac{\ell( \ell+1)}{r^2}-\frac{6M}{r^3}\right),\\
	V_{12}&=f\frac{96 \pi  \alpha M}{r^5},\qquad V_{21}=f\frac{6 \alpha M (\ell+2)!}{r^5(\ell-2)!},\label{eq:crosspot}\\
	V_{22}&=f\op[\frac{\ell(\ell+1)}{r^2}\op(1+\frac{576\pi M^2\alpha^2}{r^6}\cl)+\frac{2M}{r^3}+\mu^2\cl].\label{eq:v22}
\end{align}
To compute the modes of the spacetime, we separate the time dependence by ${\bf \Psi}(t,r)={\bf \Psi}(r) e^{-i\omega t}$ and supplement the system of equations with proper boundary conditions. These boundary conditions usually represent purely outgoing waves at infinity and ingoing waves at the horizon, which are referred to as QNMs~\cite{Berti:2009kk}. However, because of the mass term, one can also have QB states in which the scalar field is suppressed at large distances. Note that, due to the form of $V_{11}$, the gravitational field is not suppressed, behaving as an outgoing wave at infinity. Here, we will explore QB states for the scalar fields in dCS.

Near the horizon, we require purely ingoing boundary condition. We have
\be
{\bf \Psi}(r_\star \to-\infty)\approx e^{-i\omega r_\star}\sum_{i=0}^{N}(r-2M)^i\{B_{g,i},B_{\Theta,i}\} ,
\label{eq:expanhori}
\ee
where the coefficients $B_{g,i}$ and $B_{\Theta,i}$ are obtained by expanding the differential equations in powers of $(r-2M)$ near the horizon, solving for coefficients iteratively up to the order of ${\cal O}[(r-2M)^N]$. 
The boundary condition at the horizon, given by Eq.~\eqref{eq:expanhori}, depends only on two constants $(B_{g,0},B_{\Theta,0})$. 
We can, therefore, obtain numerically two independent ingoing solutions at the horizon, namely, $\bf \Psi_{1}^{(-)}$ and $\bf \Psi_{2}^{(-)}$, by choosing $(B_{g,0},B_{\Theta,0})=(1,0)$ and $(0,1)$ and integrating Eq.~\eqref{eq:perturbation} outwards. A generic solution of Eq.~\eqref{eq:perturbation} that is ingoing at the event horizon can be obtained by
\be
{\bf \Psi}_{-}=\beta_{1}^{(-)}{\bf \Psi_{1}^{(-)}}+\beta_{2}^{(-)}{\bf \Psi_{2}^{(-)}}.
\label{eq:qnm1}
\ee

At large distances, the boundary condition is more involved. We have to deal with two different asymptotic forms in order to properly describe the boundary conditions at infinity (see Appendix B of Ref.~\cite{Cardoso:2009pk} for a different approach). The first boundary condition can be obtained by making
\be
{\bf \Psi}(r_\star \to\infty)\sim r^{\mp\nu}e^{\pm k r_{\star}}\sum_{i=0}^N \op\{\frac{A_{g,i}}{r^{i+5}},\frac{A_{\Theta,i}}{r^i} \cl\},
\label{eq:binf1}
\ee
where $k=\sqrt{\mu^2-\omega^2}$, and $\nu=M\mu^2/k$. The upper signal indicates the QNM condition (${\rm Re}(k)>0$), and the bottom indicates the QB condition (${\rm Im}(k)<0$)~\cite{Rosa:2011my,Dolan:2007mj}. The second boundary condition is given by
\be
{\bf \Psi}(r_\star \to\infty)\sim e^{i\omega r_{\star}}\sum_{i=0}^N \op\{\frac{A_{g,i}}{r^{i}},\frac{A_{\Theta,i}}{r^{i+5}} \cl\}.
\label{eq:binf2}
\ee
It is worth noticing that the appearance of a damped term in the gravitational part of Eq.~\eqref{eq:binf1} 
and an outgoing wave term in the scalar part of Eq.~\eqref{eq:binf2} is due to the coupling between the two fields. 
In fact, this is the reason why the first contribution of these is $\propto r^{-5}$ [see the potentials $V_{12}$ and $V_{21}$ 
in Eq.~\eqref{eq:crosspot}]. Inserting Eqs.~\eqref{eq:binf1} and~\eqref{eq:binf2} into Eq.~\eqref{eq:perturbation} and expanding in powers of $r^{-1}$ leads to recurrence relations for the coefficients. In the end, these recurrence relations can be solved for the coefficients as functions	 of $A_{\Theta,0}$ [in the case of Eq.~\eqref{eq:binf1}] and of $A_{g,0}$ [in the case of Eq.~\eqref{eq:binf2}]. Therefore, as it happens for the boundary conditions at the horizon, we can construct two independent solutions integrating the differential Eq.~\eqref{eq:perturbation} from the numerical infinity, using $(A_{g,0},A_{\Theta,0})=(1,0)$ and $(0,1)$, namely, $\bf \Psi_1^{(+)}$ and $\bf \Psi_2^{(+)}$, respectively. Hence, the general solution satisfying the proper boundary conditions at infinity is
\be
{\bf \Psi}_+=\beta_1^{(+)}{\bf \Psi_1^{(+)}}+\beta_2^{(+)}{\bf \Psi_2^{(+)}}.
\label{eq:qnm2}
\ee

QB states can be found by requiring both boundary conditions at infinity and at the horizon to be satisfied. This means that, from Eqs.~\eqref{eq:qnm1} and~\eqref{eq:qnm2},
\begin{align}
{{\bf \Psi}_-}&={{\bf \Psi}_+},\label{eq:cond1}\\
\frac{d}{dr_\star}{{\bf \Psi}_-}&=\frac{d}{dr_\star}{{\bf \Psi}_+},\label{eq:cond2}
\end{align}
holds when $\omega$ is the frequency of the QB state. In practice, Eqs.~\eqref{eq:cond1} and \eqref{eq:cond2} are computed at some intermediate point, say $r=r_m$. As the system is linear, we can set one of the $\beta$ coefficients to unity and use three of the components of Eqs.~\eqref{eq:cond1} and \eqref{eq:cond2} to solve for the remaining $\beta$. The last component can be solved to find the mode frequencies $\omega$. Equivalently, we can impose that the determinant
\be
\det({\bf \Psi_{1}^{(-)}},{\bf \Psi_{2}^{(-)}},{\bf \Psi_{1}^{(+)}},{\bf \Psi_{2}^{(+)}})
\label{eq:det}
\ee 
vanishes when the frequency is the eigenvalue of the system, generating complex values for the frequency, namely, $\omega=\omega_R+i\,\omega_I$, with $(\omega_R,\omega_I)$ being real values.

The method described above is usually referred to as direct integration method (DI)~\cite{Chandrasekhar:1975zza,Pani:2013pma,Macedo:2016wgh}, and it has been applied in the past to other classes of quadratic gravity~\cite{Blazquez-Salcedo:2016enn,Blazquez-Salcedo:2016yka}. 
Another method to compute the modes and frequencies that is more efficient is the continued fraction (CF) method~\cite{Leaver:1985ax,Nollert:1993zz}. The CF method relies on finding an appropriate expression to represent the fields such that it satisfies the proper boundary conditions of the problem (see, e.g., Refs.~\cite{Pani:2013pma,Macedo:2016wgh}). 
For coupled systems, such as the one presented in this paper, the situations are more complex because we have to consider a matrix-valued recurrence relation (see Refs.~\cite{Cardoso:2009pk,Pani:2013pma} for instance). For our specific problem, we also have to find an expression that handles both with the behavior of outgoing waves and damped for the perturbations as presented in Eqs.~\eqref{eq:binf1} and~\eqref{eq:binf2}. Therefore, a more robust method---such as the CF---to find the modes is beyond the scope of this paper. Notwithstanding, the DI method usually handles QB states computations well, as in the case of the parameter space explored in this paper.

\subsection{Monopolar and dipolar modes}

Differently from GR, axial perturbations in dCS gravity pre\-sent monopolar ($\ell=0$) and dipolar ($\ell=1$) radiative modes, linked with the scalar degree of freedom of the theory. One can show that the these modes can be described by the same equation, given by
\be
\left(\frac{d}{dr_\star}-V_{22}\right)\Theta=0,
\label{eq:dec}
\ee 
where $V_{22}$ is given by Eq.~\eqref{eq:v22}. Because the system reduces to a single second-order differential equation, 
it is much simpler to analyze, compared to the $\ell>1$ modes. Note that for $\ell=0$, the modes 
have a behavior that is precisely the same as that of a massive scalar field in the Schwarzschild background because the term proportional 
to $\alpha$ vanishes. On the other hand, the $\ell=1$ mode does have an additional term proportional to the coupling parameter. 
The $\ell=1$ mode is especially important because it is this mode that presents the most prominent superradiant instabilitiy 
in rotating BH spacetimes~\cite{Dolan:2012yt}. Therefore, one can analyze if the coupling makes the mode less susceptible 
to this instability by looking into the influence of the coupling into the imaginary part.

One may be tempted to adopt Eq.~\eqref{eq:dec} as an approximation scheme for higher multipoles, which would be the case for a Dudley-Finley-like scheme~\cite{Dudley:1977zz}. However, we verified that, although it reduces to the GR case in the limit $\alpha\to 0$, Eq.~~\eqref{eq:dec} \textit{does not} reproduce the behavior of the modes computed using the full set of equations~\eqref{eq:perturbation}. This reflects the importance of the gravitational feedback into the scalar QB modes.

\subsection{Breit-Wigner method for the bound states}
Another method that we can exploit is the Breit-Wigner resonant method~\cite{1991RSPSA.434..449C,Berti:2009wx,Pani:2013pma,Macedo:2018yoi}. This method works well for modes with high-quality factors, i.e., 
$|\omega_R/\omega_I|\gg 1.$
Essentially, we have that the determinant given by Eq.~\eqref{eq:det} near the QB frequencies can be expanded as~\cite{Pani:2013pma}
\be
|\det({\bf S})|^2\propto (\omega-\omega_R)^2+\omega_I^2.
\label{eq:bw}
\ee
The modes can be found by sweeping $\omega$ through real values, finding the above determinant, and fitting the modes to Eq.~\eqref{eq:bw}. As QB states are usually high-quality factor modes, we can use the Breit-Wigner method as a check for the modes computed using the full direction integration method.

\subsection{Time evolution of initial data}

In order to understand the influence of the QB states into gravitational wave signals, we also perform a time evolution of the system, similarly to the one presented in Refs.~\cite{Gundlach:1993tp,Molina:2010fb}. We write the system~\eqref{eq:perturbation} using the light-cone coordinates, $u=r_\star-t$ and $v=r_\star+t$, obtaining
\be
4\frac{\partial^2}{\partial u\,\partial v}{\bf \Psi}=-{\bf V}{\bf \Psi}.
\label{eq:systemp}
\ee
We solve the system~\eqref{eq:systemp} with Gaussian initial conditions, i.e.,
\begin{align}
{\bf \Psi}(0,v)&=\left[
\begin{array}{l}
	A_1	 \exp\left(-\frac{(v-v_{c1})^2}{2 	\sigma_1}\right)\\
	A_2	 \exp\left(-\frac{(v-v_{c2})^2}{2 	\sigma_2}\right)	
\end{array}\right],~{\rm and}~\label{eq:init1}\\
{\bf \Psi}(u,0)&=\left[
\begin{array}{l}
0\\
0	
\end{array}\right].
\label{eq:init2}
\end{align}
In general, it is expected that the results, after a transient regime, do not depend on the choice of initial data~\cite{Nollert:1993zz}. As such, for our purpose, it is sufficient to consider only the initial Gaussian data of the type given by Eqs.~\eqref{eq:init1} and~\eqref{eq:init2}. 

For massive fields, time evolutions shed light in many directions. One can study the frequency of the QB states by analyzing the late-time signal through a fre\-quen\-cy-fil\-ter technique~\cite{Dolan:2012yt}. One can also look into the influence of the scalar field on the
 gravitational signals, which can be studied, for instance, through only gravitational initial data [i.e., setting $A_2=0$ in Eq.~\eqref{eq:init1}]. Because of the coupling, after an initial time, the scalar field will have an influence in the gravitational wave signal (and \textit{vice versa}). Therefore, the mass of the scalar field will have an influence on the power-law tail. For the massless case, the power-law tail behavior is given by $\propto t^{-(2\ell+3)}$. For the massive case, it is given by~\cite{Koyama:2001ee,Koyama:2001qw,Burko:2004jn,Witek:2012tr}
\begin{equation}
{\bf \Psi}\sim t^{p} \sin(\omega_c t),
\label{eq:latetime}
\end{equation}
with $p=-(\ell+3/2)$ at intermediate times and $p=-5/6$ at very late times, and $\omega_c$ spans over values close to the scalar field's mass, typically related to the QB state frequencies. This gives us possible distinctive signatures of alternative theories with extra massive degrees of freedom.

\section{Numerical results}\label{sec:results}
 
Here, we present our numerical results, parametrizing them by $\zeta=16\pi\alpha^2/M^4$. In order to validate our numerical code, we computed the QNMs in the massless case and performed time evolutions in the GR limit. For the massless case, we compare our the results with the ones presented in Ref.~\cite{Molina:2010fb}, and the time evolution and massive modes with~\cite{Burko:2004jn,Dolan:2007mj}. Our results are in	 excellent agreement with the ones presented in the literature.

\subsection{Quasibound states}

\begin{figure}%
\includegraphics[width=0.95\columnwidth]{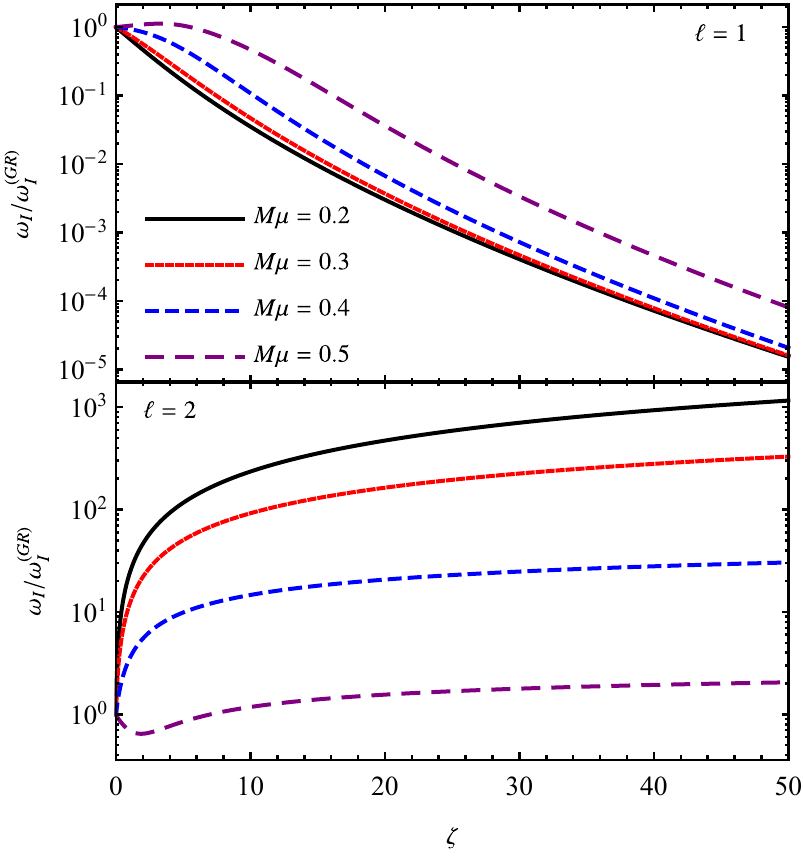}%
\caption{Imaginary part of the fundamental QB frequencies, normalized by the GR value, as a function of the coupling constant $\zeta$. \textit{Top panel}: For the $\ell=1$ mode we see that, in the mass range studied here, the  mode decreases in absolute value for high values of the coupling $\zeta$, meaning that it is becoming less stable. \textit{Bottom panel:} The $\ell=2$ mode behaves differently from the $\ell=1$ mode, increasing in absolute value, becoming more stable. For both $\ell=1$ and $\ell=2$, these effects are attenuated for higher values of the mass.}%
\label{fig:mode1}%
\end{figure}
We apply the DI procedure described in Sec.~\ref{sec:pert}. We focus on the dipolar and quadrupolar modes, but we also verify that the results for $\ell=3$ are qualitatively similar to the $\ell=2$ ones. We recall that the monopolar mode ($\ell=0$) is precisely the same as that in GR. The dipolar case is important in order study possible implications of the coupling constant into the superradiant instability, once rotation is considered. The quadrupolar mode can show possible signatures in gravitational wave observations.

\begin{figure*}%
\includegraphics[width=\columnwidth]{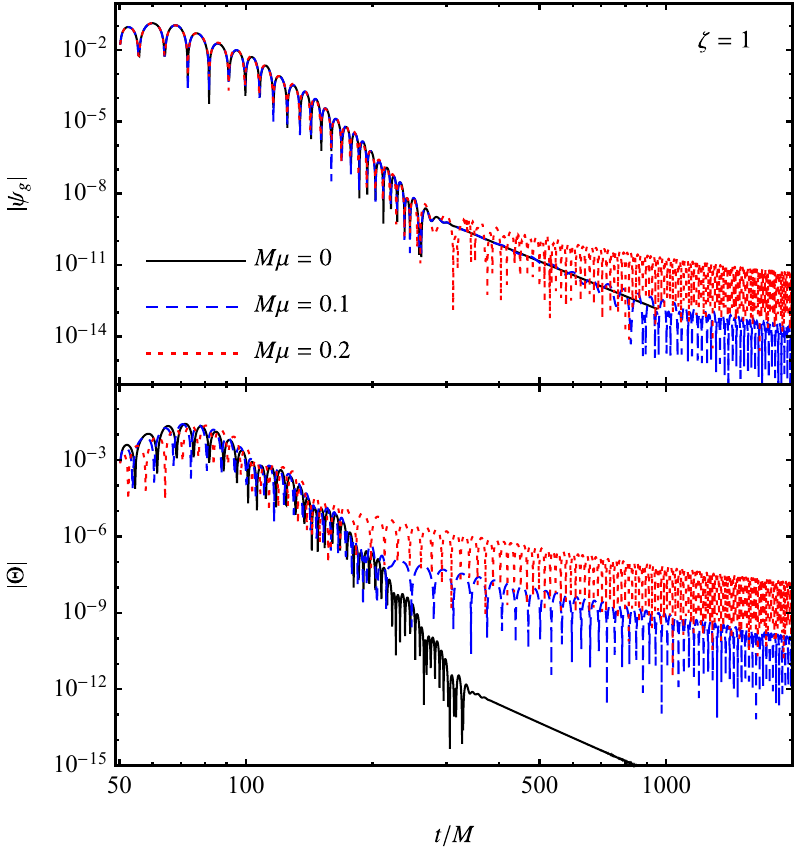}\includegraphics[width=\columnwidth]{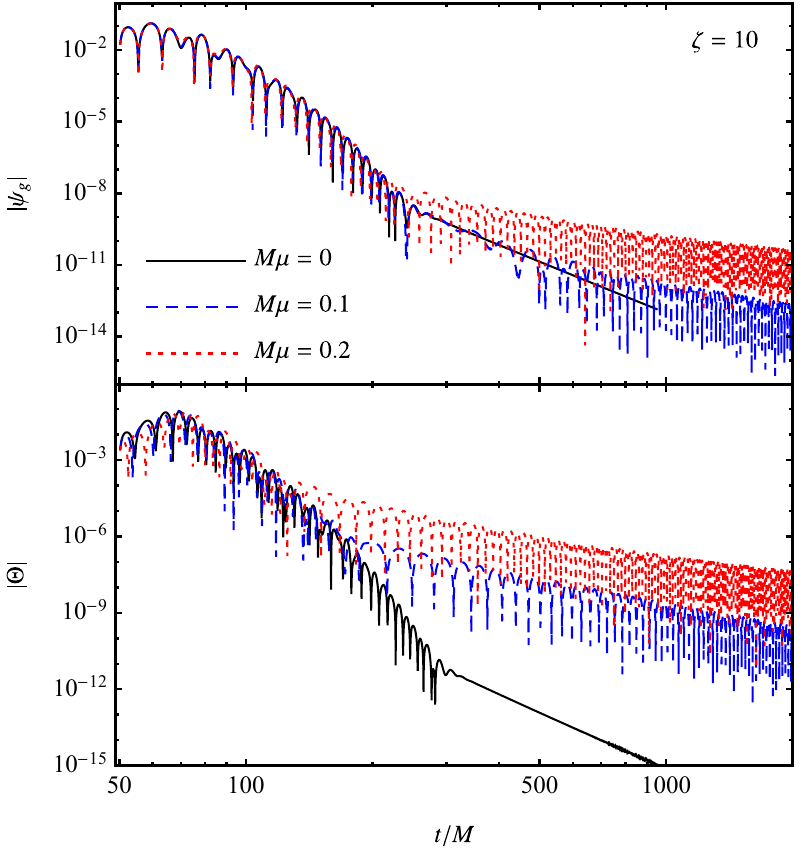}%
\caption{Time evolution of an initial quadrupolar ($\ell=2$) gravitational Gaussian profile for different scalar masses, $\zeta=1$ (left panels) and $\zeta=10$ (right panels). \textit{Top panel:} Gravitational perturbation is initially insensitive to the scalar influence, but later on, it becomes contaminated by it, presenting an oscillatory behavior. \textit{Bottom panel:} Scalar perturbation is initially smaller than the gravitational perturbation, becoming larger after some time. }%
\label{fig:time1}%
\end{figure*}

\begin{figure*}%
\includegraphics[width=\columnwidth]{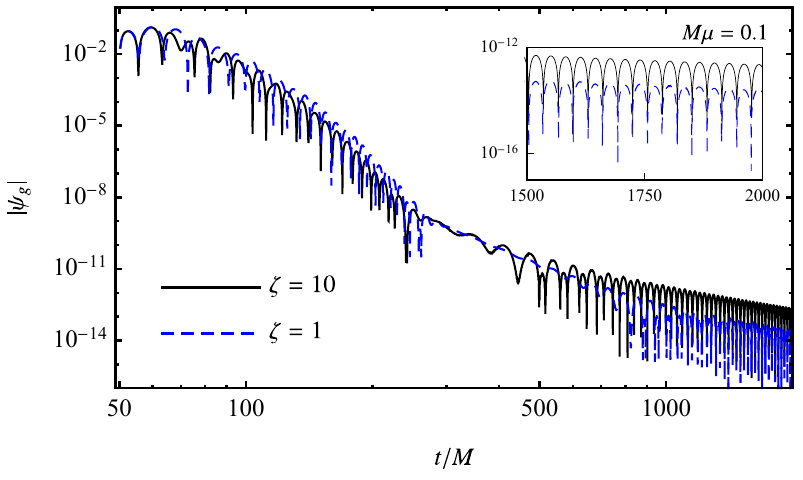}\includegraphics[width=\columnwidth]{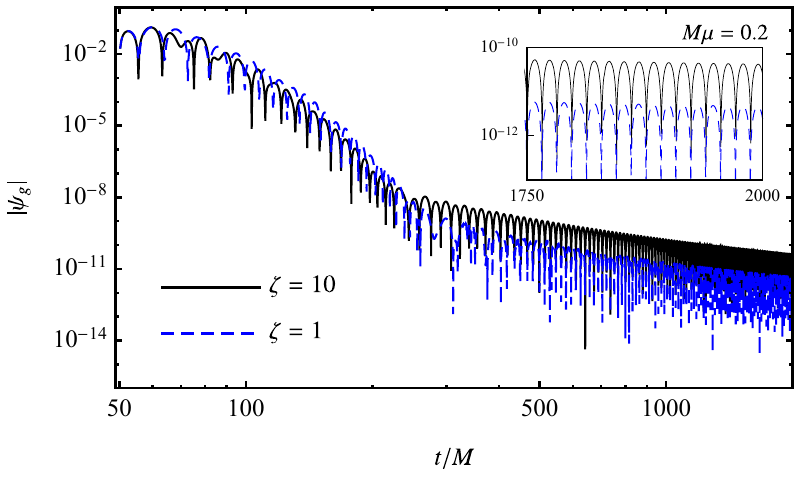}%
\caption{Time evolution of gravitational wave packets considering different values for the coupling $\zeta$, $M\mu=0.1$ (left panels) and $M\mu=0.2$ (right panels). We see that the characteristic oscillation appears earlier for higher values of the coupling $\zeta$, but the frequency of these oscillations is basically the same (see the inset).}%
\label{fig:time2}%
\end{figure*}
In Fig.~\ref{fig:mode1} we show the imaginary part of the $\ell=1$ and $\ell=2$ fundamental QB frequency, 
normalized by the GR value ($\omega_{I}^{(GR)}$), for different values of the mass $M\mu$. The behavior of the dipolar modes is very different from that of the quadrupolar one. 
In the top panel of Fig.~\ref{fig:mode1}, we see that the dipolar mode becomes less stable as the coupling $\zeta$ grows, meaning that the imaginary part decreases in absolute value. This implies that these modes live even longer than they do in GR, making them more susceptible to the superradiant instability once rotation is considered. The opposite happens for the imaginary part of the $\ell=2$ mode (bottom panel of Fig.~\ref{fig:mode1}).
We see that the imaginary part increases in absolute value, becoming more stable and, therefore, more damped. We confirm these results by estimating the mode frequencies looking into the Breit-Wigner expression presented in Sec.~\ref{sec:pert}, comparing with the DI result.

We note that the behavior described above switches when considering $M\mu\sim0.5$ for small couplings. As we can see in Fig.~\ref{fig:mode1}, there is a maximum in the $\ell=1$ case and a minimum for the $\ell=2$ case. Nonetheless, for high values of the coupling, the behavior of the modes is the same as that described above, i.e., $\ell=1$ modes gets less stable and $\ell=2$ more stable.

For the real part of the modes, we find that in all cases it is basically insensitive to changes in the coupling $\zeta$, the differences being $<0.1\%$ from the GR modes in the range considered here. Hence, in the massive dCS gravity, the oscillation of the perturbations remains basically the same, but the damping time depends on the coupling.

In all cases described here, we performed a numerical search for unstable QB modes, finding none. Our numerical results further corroborate the conclusion that the Schwarzschild spacetime in massive dCS gravity is stable. Note that this conclusion is supported from studies of another independent method, the $S$-deformation method~\cite{Kimura:2018nxk}.

\subsection{Time evolution of gravitational pulses}


Here, focus on numerical evolution of an initially pure gravitational Gaussian packet in order to understand the influence of the scalar sector on gravitational waves, considering $A_2=0$ in Eq.~\eqref{eq:init1}. The Gaussian packet is centered at $v_{c1}=10M$ and has a width $\sigma=M$, and the wave functions are extracted at $r_\star=50M$.

In Fig.~\ref{fig:time1} we show the result of a time evolution considering different values for the scalar field mass, considering $\zeta=1$ (left panels) and $\zeta=10$ (right panels). We can see that the initial gravitational response for all cases is the same. After some time, the gravitational field becomes contaminated with the scalar field, which impacts the behavior of the late-time tails not only in the power-law profile, but also by oscillations. The frequency of these oscillations is predominantly given by the QB frequencies, similarly to what happens in the GR case~\cite{Burko:2004jn} (see also Ref.~\cite{Degollado:2014vsa}). The scalar field perturbation is initially smaller than the gravitational one, as expected because the pulse is initially only gravitational. However, the scalar perturbation grows in time, becoming larger at some intermediate times depending on  $M\mu$  and $\zeta$.

\begin{figure}%
\includegraphics[width=\columnwidth]{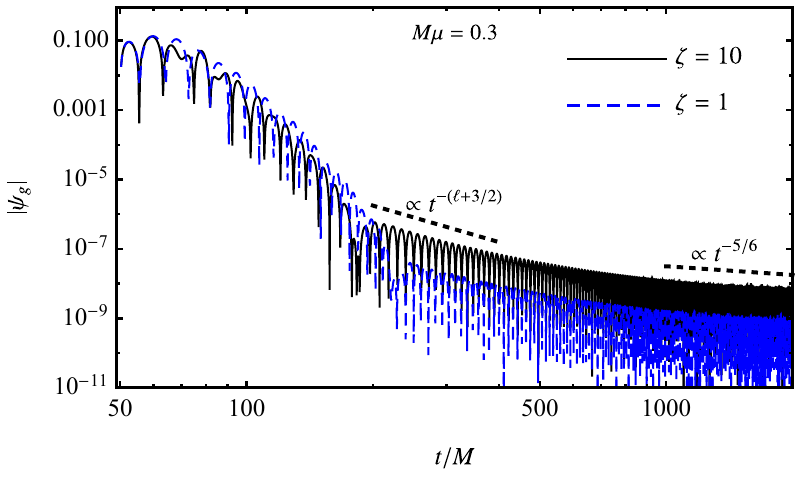}%
\caption{Time evolution of gravitational wave packets for $M\mu=0.3$, considering $\zeta=1$ and $\zeta=10$. In this plot, we can see that the intermediate and very late time behavior of the gravitational perturbation is similar to the massive case in pure GR, given by Eq.~\eqref{eq:latetime}.}%
\label{fig:tail}%
\end{figure}
To see the role of the coupling constant on the gravitational signal, in Fig.~\ref{fig:time2} we plot the results for the time evolution with $M\mu=0.1$ and $M\mu=0.2$, considering $\zeta=1$ and $\zeta=10$. 
For $\zeta=10$ the scalar field has an impact on the gravitational perturbations at earlier times than in the $\zeta=1$ case. 
Additionally, in the late-time regime, the amplitude of the gravitational wave is larger in the $\zeta=10$ than it is for $\zeta=1$, demonstrating a clear influence of the cross terms in the late-time behavior. Also, we can see that the oscillation of the late field is the same for the two cases, which corroborates our result that the real part of the modes is almost insensitive to changes in $\zeta$.

In Fig.~\ref{fig:tail} we plot the results for the time evolution with $M\mu=0.3$, considering $\zeta=1$ and $\zeta=10$. We can readily see that the power law for the gravitational perturbation is not the same as that in GR. Instead, we see that the behavior fits the ones given by Eq.~\eqref{eq:latetime}, which shows further imprints of the scalar field's mass onto the gravitational sector of the perturbations.

\section{Discussion and conclusion}\label{sec:conclusion}

Perturbation of BH spacetimes is a timely topic, helping us to understand the dynamics of these objects with a linear system of equations. In this work, we studied axial perturbations of static, spherically symmetric BHs in dCS gravity with a massive scalar field. Our work extends previous works in the literature, analyzing the behavior of QB states, which are long-living modes (partially) damped by the mass term. We found that the real part of the modes is insensitive to the coupling constant, and that the dipolar mode becomes less stable and the quadrupole more stable, as the coupling constant increases. We also analyzed the influence of the mass term into gravitational signals propagating in the BH spacetime, through the study of time evolutions of gravitational Gaussian packets. We showed that the scalar field mass has a direct influence on the late-time behavior of the signals.

Considering the QB states, our results show that BHs in massive dCS gravity can be, in principle, 
more susceptible to the superradiant instabilities of rotating black holes when compared to the GR counterpart. This susceptibility 
will depend on how the perturbations are coupled when rotation is considered. The results presented here give an indication of QB mode dependence on the coupling constant. A further study considering rotating black holes is needed to present the whole picture in this scenario and, therefore, is a logical extension of the present paper.

For the time evolution of gravitational perturbations, our results show that the theory presents clear signatures at late times, which could be potentially observed once we can access the gravitational tail. Additionally, this late-time behavior could be a signature of theories that are coupled with massive fields, not being exclusive to massive dCS gravity. For instance, within GR, if one considers the  gravitational radiation triggered by the interaction of massive scalar field packets with black holes, the power-law tail will oscillate with a characteristic frequency that depends essentially on the QB frequency of the scalar field~\cite{Degollado:2014vsa}, similarly to what we described here. Clearly, because the scalar field in dCS is directly coupled with the gravitational perturbations, the oscillation is more pronounced, as it could possibly be the case of other similar theories of gravity.

\section*{Acknowledgments}
It is a pleasure to thank Hector O. Silva, Leandro A. Oliveira, Luís C. B. Crispino, Paolo Pani, and Vitor Cardoso for useful comments and discussions. I would like to thank Conselho Nacional de Desenvolvimento Científico e Tecnológico
(CNPq), Coordenação de Aperfeiçoamento de Pessoal
de Nível Superior (CA\-PES), and Fundação Amazônia de Amparo a Estudos e Pesquisas (FAPESPA), from Brazil, for partial financial
support.

  \bibliography{biblio}

\begin{thebibliography}{51}%
\makeatletter
\providecommand \@ifxundefined [1]{%
 \@ifx{#1\undefined}
}%
\providecommand \@ifnum [1]{%
 \ifnum #1\expandafter \@firstoftwo
 \else \expandafter \@secondoftwo
 \fi
}%
\providecommand \@ifx [1]{%
 \ifx #1\expandafter \@firstoftwo
 \else \expandafter \@secondoftwo
 \fi
}%
\providecommand \natexlab [1]{#1}%
\providecommand \enquote  [1]{``#1''}%
\providecommand \bibnamefont  [1]{#1}%
\providecommand \bibfnamefont [1]{#1}%
\providecommand \citenamefont [1]{#1}%
\providecommand \href@noop [0]{\@secondoftwo}%
\providecommand \href [0]{\begingroup \@sanitize@url \@href}%
\providecommand \@href[1]{\@@startlink{#1}\@@href}%
\providecommand \@@href[1]{\endgroup#1\@@endlink}%
\providecommand \@sanitize@url [0]{\catcode `\\12\catcode `\$12\catcode
  `\&12\catcode `\#12\catcode `\^12\catcode `\_12\catcode `\%12\relax}%
\providecommand \@@startlink[1]{}%
\providecommand \@@endlink[0]{}%
\providecommand \url  [0]{\begingroup\@sanitize@url \@url }%
\providecommand \@url [1]{\endgroup\@href {#1}{\urlprefix }}%
\providecommand \urlprefix  [0]{URL }%
\providecommand \Eprint [0]{\href }%
\providecommand \doibase [0]{http://dx.doi.org/}%
\providecommand \selectlanguage [0]{\@gobble}%
\providecommand \bibinfo  [0]{\@secondoftwo}%
\providecommand \bibfield  [0]{\@secondoftwo}%
\providecommand \translation [1]{[#1]}%
\providecommand \BibitemOpen [0]{}%
\providecommand \bibitemStop [0]{}%
\providecommand \bibitemNoStop [0]{.\EOS\space}%
\providecommand \EOS [0]{\spacefactor3000\relax}%
\providecommand \BibitemShut  [1]{\csname bibitem#1\endcsname}%
\let\auto@bib@innerbib\@empty
\bibitem [{\citenamefont {Abbott}\ \emph
  {et~al.}(2016{\natexlab{a}})\citenamefont {Abbott} \emph
  {et~al.}}]{Abbott:2016blz}%
  \BibitemOpen
  \bibfield  {author} {\bibinfo {author} {\bibfnamefont {B.~P.}\ \bibnamefont
  {Abbott}} \emph {et~al.} (\bibinfo {collaboration} {Virgo, LIGO
  Scientific}),\ }\href {\doibase 10.1103/PhysRevLett.116.061102} {\bibfield
  {journal} {\bibinfo  {journal} {Phys. Rev. Lett.}\ }\textbf {\bibinfo
  {volume} {116}},\ \bibinfo {pages} {061102} (\bibinfo {year}
  {2016}{\natexlab{a}})},\ \Eprint {http://arxiv.org/abs/1602.03837}
  {arXiv:1602.03837 [gr-qc]} \BibitemShut {NoStop}%
\bibitem [{\citenamefont {Abbott}\ \emph
  {et~al.}(2016{\natexlab{b}})\citenamefont {Abbott} \emph
  {et~al.}}]{Abbott:2016nmj}%
  \BibitemOpen
  \bibfield  {author} {\bibinfo {author} {\bibfnamefont {B.~P.}\ \bibnamefont
  {Abbott}} \emph {et~al.} (\bibinfo {collaboration} {Virgo, LIGO
  Scientific}),\ }\href {\doibase 10.1103/PhysRevLett.116.241103} {\bibfield
  {journal} {\bibinfo  {journal} {Phys. Rev. Lett.}\ }\textbf {\bibinfo
  {volume} {116}},\ \bibinfo {pages} {241103} (\bibinfo {year}
  {2016}{\natexlab{b}})},\ \Eprint {http://arxiv.org/abs/1606.04855}
  {arXiv:1606.04855 [gr-qc]} \BibitemShut {NoStop}%
\bibitem [{\citenamefont {Abbott}\ \emph
  {et~al.}(2017{\natexlab{a}})\citenamefont {Abbott} \emph
  {et~al.}}]{Abbott:2017oio}%
  \BibitemOpen
  \bibfield  {author} {\bibinfo {author} {\bibfnamefont {B.~P.}\ \bibnamefont
  {Abbott}} \emph {et~al.} (\bibinfo {collaboration} {Virgo, LIGO
  Scientific}),\ }\href {\doibase 10.1103/PhysRevLett.119.141101} {\bibfield
  {journal} {\bibinfo  {journal} {Phys. Rev. Lett.}\ }\textbf {\bibinfo
  {volume} {119}},\ \bibinfo {pages} {141101} (\bibinfo {year}
  {2017}{\natexlab{a}})},\ \Eprint {http://arxiv.org/abs/1709.09660}
  {arXiv:1709.09660 [gr-qc]} \BibitemShut {NoStop}%
\bibitem [{\citenamefont {Abbott}\ \emph
  {et~al.}(2017{\natexlab{b}})\citenamefont {Abbott} \emph
  {et~al.}}]{Abbott:2017vtc}%
  \BibitemOpen
  \bibfield  {author} {\bibinfo {author} {\bibfnamefont {B.~P.}\ \bibnamefont
  {Abbott}} \emph {et~al.} (\bibinfo {collaboration} {VIRGO, LIGO
  Scientific}),\ }\href {\doibase 10.1103/PhysRevLett.118.221101} {\bibfield
  {journal} {\bibinfo  {journal} {Phys. Rev. Lett.}\ }\textbf {\bibinfo
  {volume} {118}},\ \bibinfo {pages} {221101} (\bibinfo {year}
  {2017}{\natexlab{b}})},\ \Eprint {http://arxiv.org/abs/1706.01812}
  {arXiv:1706.01812 [gr-qc]} \BibitemShut {NoStop}%
\bibitem [{\citenamefont {Abbott}\ \emph
  {et~al.}(2017{\natexlab{c}})\citenamefont {Abbott} \emph
  {et~al.}}]{TheLIGOScientific:2017qsa}%
  \BibitemOpen
  \bibfield  {author} {\bibinfo {author} {\bibfnamefont {B.}~\bibnamefont
  {Abbott}} \emph {et~al.} (\bibinfo {collaboration} {Virgo, LIGO
  Scientific}),\ }\href {\doibase 10.1103/PhysRevLett.119.161101} {\bibfield
  {journal} {\bibinfo  {journal} {Phys. Rev. Lett.}\ }\textbf {\bibinfo
  {volume} {119}},\ \bibinfo {pages} {161101} (\bibinfo {year}
  {2017}{\natexlab{c}})},\ \Eprint {http://arxiv.org/abs/1710.05832}
  {arXiv:1710.05832 [gr-qc]} \BibitemShut {NoStop}%
\bibitem [{\citenamefont {Abbott}\ \emph
  {et~al.}(2017{\natexlab{d}})\citenamefont {Abbott} \emph
  {et~al.}}]{Abbott:2017gyy}%
  \BibitemOpen
  \bibfield  {author} {\bibinfo {author} {\bibfnamefont {B.~P.}\ \bibnamefont
  {Abbott}} \emph {et~al.} (\bibinfo {collaboration} {Virgo, LIGO
  Scientific}),\ }\href {\doibase 10.3847/2041-8213/aa9f0c} {\bibfield
  {journal} {\bibinfo  {journal} {Astrophys. J.}\ }\textbf {\bibinfo {volume}
  {851}},\ \bibinfo {pages} {L35} (\bibinfo {year} {2017}{\natexlab{d}})},\
  \Eprint {http://arxiv.org/abs/1711.05578} {arXiv:1711.05578 [astro-ph.HE]}
  \BibitemShut {NoStop}%
\bibitem [{\citenamefont {Abbott}\ \emph
  {et~al.}(2016{\natexlab{c}})\citenamefont {Abbott} \emph
  {et~al.}}]{TheLIGOScientific:2016src}%
  \BibitemOpen
  \bibfield  {author} {\bibinfo {author} {\bibfnamefont {B.~P.}\ \bibnamefont
  {Abbott}} \emph {et~al.} (\bibinfo {collaboration} {Virgo, LIGO
  Scientific}),\ }\href {\doibase 10.1103/PhysRevLett.116.221101} {\bibfield
  {journal} {\bibinfo  {journal} {Phys. Rev. Lett.}\ }\textbf {\bibinfo
  {volume} {116}},\ \bibinfo {pages} {221101} (\bibinfo {year}
  {2016}{\natexlab{c}})},\ \Eprint {http://arxiv.org/abs/1602.03841}
  {arXiv:1602.03841 [gr-qc]} \BibitemShut {NoStop}%
\bibitem [{\citenamefont {Konoplya}\ and\ \citenamefont
  {Zhidenko}(2016)}]{Konoplya:2016pmh}%
  \BibitemOpen
  \bibfield  {author} {\bibinfo {author} {\bibfnamefont {R.}~\bibnamefont
  {Konoplya}}\ and\ \bibinfo {author} {\bibfnamefont {A.}~\bibnamefont
  {Zhidenko}},\ }\href {\doibase 10.1016/j.physletb.2016.03.044} {\bibfield
  {journal} {\bibinfo  {journal} {Phys. Lett.}\ }\textbf {\bibinfo {volume}
  {B756}},\ \bibinfo {pages} {350} (\bibinfo {year} {2016})},\ \Eprint
  {http://arxiv.org/abs/1602.04738} {arXiv:1602.04738 [gr-qc]} \BibitemShut
  {NoStop}%
\bibitem [{\citenamefont {Yunes}\ \emph {et~al.}(2016)\citenamefont {Yunes},
  \citenamefont {Yagi},\ and\ \citenamefont {Pretorius}}]{Yunes:2016jcc}%
  \BibitemOpen
  \bibfield  {author} {\bibinfo {author} {\bibfnamefont {N.}~\bibnamefont
  {Yunes}}, \bibinfo {author} {\bibfnamefont {K.}~\bibnamefont {Yagi}}, \ and\
  \bibinfo {author} {\bibfnamefont {F.}~\bibnamefont {Pretorius}},\ }\href
  {\doibase 10.1103/PhysRevD.94.084002} {\bibfield  {journal} {\bibinfo
  {journal} {Phys. Rev.}\ }\textbf {\bibinfo {volume} {D94}},\ \bibinfo {pages}
  {084002} (\bibinfo {year} {2016})},\ \Eprint
  {http://arxiv.org/abs/1603.08955} {arXiv:1603.08955 [gr-qc]} \BibitemShut
  {NoStop}%
\bibitem [{\citenamefont {Berti}\ \emph {et~al.}(2015)\citenamefont {Berti}
  \emph {et~al.}}]{Berti:2015itd}%
  \BibitemOpen
  \bibfield  {author} {\bibinfo {author} {\bibfnamefont {E.}~\bibnamefont
  {Berti}} \emph {et~al.},\ }\href {\doibase 10.1088/0264-9381/32/24/243001}
  {\bibfield  {journal} {\bibinfo  {journal} {Class. Quant. Grav.}\ }\textbf
  {\bibinfo {volume} {32}},\ \bibinfo {pages} {243001} (\bibinfo {year}
  {2015})},\ \Eprint {http://arxiv.org/abs/1501.07274} {arXiv:1501.07274
  [gr-qc]} \BibitemShut {NoStop}%
\bibitem [{\citenamefont {Cardoso}\ and\ \citenamefont
  {Gualtieri}(2016)}]{Cardoso:2016ryw}%
  \BibitemOpen
  \bibfield  {author} {\bibinfo {author} {\bibfnamefont {V.}~\bibnamefont
  {Cardoso}}\ and\ \bibinfo {author} {\bibfnamefont {L.}~\bibnamefont
  {Gualtieri}},\ }\href {\doibase 10.1088/0264-9381/33/17/174001} {\bibfield
  {journal} {\bibinfo  {journal} {Class. Quant. Grav.}\ }\textbf {\bibinfo
  {volume} {33}},\ \bibinfo {pages} {174001} (\bibinfo {year} {2016})},\
  \Eprint {http://arxiv.org/abs/1607.03133} {arXiv:1607.03133 [gr-qc]}
  \BibitemShut {NoStop}%
\bibitem [{\citenamefont {Berti}\ \emph
  {et~al.}(2018{\natexlab{a}})\citenamefont {Berti}, \citenamefont {Yagi},\
  and\ \citenamefont {Yunes}}]{Berti:2018cxi}%
  \BibitemOpen
  \bibfield  {author} {\bibinfo {author} {\bibfnamefont {E.}~\bibnamefont
  {Berti}}, \bibinfo {author} {\bibfnamefont {K.}~\bibnamefont {Yagi}}, \ and\
  \bibinfo {author} {\bibfnamefont {N.}~\bibnamefont {Yunes}},\ }\href
  {\doibase 10.1007/s10714-018-2362-8} {\bibfield  {journal} {\bibinfo
  {journal} {Gen. Rel. Grav.}\ }\textbf {\bibinfo {volume} {50}},\ \bibinfo
  {pages} {46} (\bibinfo {year} {2018}{\natexlab{a}})},\ \Eprint
  {http://arxiv.org/abs/1801.03208} {arXiv:1801.03208 [gr-qc]} \BibitemShut
  {NoStop}%
\bibitem [{\citenamefont {Berti}\ \emph
  {et~al.}(2018{\natexlab{b}})\citenamefont {Berti}, \citenamefont {Yagi},
  \citenamefont {Yang},\ and\ \citenamefont {Yunes}}]{Berti:2018vdi}%
  \BibitemOpen
  \bibfield  {author} {\bibinfo {author} {\bibfnamefont {E.}~\bibnamefont
  {Berti}}, \bibinfo {author} {\bibfnamefont {K.}~\bibnamefont {Yagi}},
  \bibinfo {author} {\bibfnamefont {H.}~\bibnamefont {Yang}}, \ and\ \bibinfo
  {author} {\bibfnamefont {N.}~\bibnamefont {Yunes}},\ }\href {\doibase
  10.1007/s10714-018-2372-6} {\bibfield  {journal} {\bibinfo  {journal} {Gen.
  Rel. Grav.}\ }\textbf {\bibinfo {volume} {50}},\ \bibinfo {pages} {49}
  (\bibinfo {year} {2018}{\natexlab{b}})},\ \Eprint
  {http://arxiv.org/abs/1801.03587} {arXiv:1801.03587 [gr-qc]} \BibitemShut
  {NoStop}%
\bibitem [{\citenamefont {Barack}\ \emph {et~al.}(2018)\citenamefont {Barack}
  \emph {et~al.}}]{Barack:2018yly}%
  \BibitemOpen
  \bibfield  {author} {\bibinfo {author} {\bibfnamefont {L.}~\bibnamefont
  {Barack}} \emph {et~al.},\ }\href@noop {} {\  (\bibinfo {year} {2018})},\
  \Eprint {http://arxiv.org/abs/1806.05195} {arXiv:1806.05195 [gr-qc]}
  \BibitemShut {NoStop}%
\bibitem [{\citenamefont {Alexander}\ and\ \citenamefont
  {Yunes}(2009)}]{Alexander:2009tp}%
  \BibitemOpen
  \bibfield  {author} {\bibinfo {author} {\bibfnamefont {S.}~\bibnamefont
  {Alexander}}\ and\ \bibinfo {author} {\bibfnamefont {N.}~\bibnamefont
  {Yunes}},\ }\href {\doibase 10.1016/j.physrep.2009.07.002} {\bibfield
  {journal} {\bibinfo  {journal} {Phys. Rept.}\ }\textbf {\bibinfo {volume}
  {480}},\ \bibinfo {pages} {1} (\bibinfo {year} {2009})},\ \Eprint
  {http://arxiv.org/abs/0907.2562} {arXiv:0907.2562 [hep-th]} \BibitemShut
  {NoStop}%
\bibitem [{\citenamefont {Jackiw}\ and\ \citenamefont
  {Pi}(1990)}]{Jackiw:1990mb}%
  \BibitemOpen
  \bibfield  {author} {\bibinfo {author} {\bibfnamefont {R.}~\bibnamefont
  {Jackiw}}\ and\ \bibinfo {author} {\bibfnamefont {S.-Y.}\ \bibnamefont
  {Pi}},\ }\href {\doibase 10.1103/PhysRevD.42.3500, 10.1103/PhysRevD.48.3929}
  {\bibfield  {journal} {\bibinfo  {journal} {Phys. Rev.}\ }\textbf {\bibinfo
  {volume} {D42}},\ \bibinfo {pages} {3500} (\bibinfo {year} {1990})},\
  \bibinfo {note} {[Erratum: Phys. Rev.D48,3929(1993)]}\BibitemShut {NoStop}%
\bibitem [{\citenamefont {Campbell}\ \emph {et~al.}(1991)\citenamefont
  {Campbell}, \citenamefont {Duncan}, \citenamefont {Kaloper},\ and\
  \citenamefont {Olive}}]{Campbell:1990fu}%
  \BibitemOpen
  \bibfield  {author} {\bibinfo {author} {\bibfnamefont {B.~A.}\ \bibnamefont
  {Campbell}}, \bibinfo {author} {\bibfnamefont {M.~J.}\ \bibnamefont
  {Duncan}}, \bibinfo {author} {\bibfnamefont {N.}~\bibnamefont {Kaloper}}, \
  and\ \bibinfo {author} {\bibfnamefont {K.~A.}\ \bibnamefont {Olive}},\ }\href
  {\doibase 10.1016/S0550-3213(05)80045-8} {\bibfield  {journal} {\bibinfo
  {journal} {Nucl. Phys.}\ }\textbf {\bibinfo {volume} {B351}},\ \bibinfo
  {pages} {778} (\bibinfo {year} {1991})}\BibitemShut {NoStop}%
\bibitem [{\citenamefont {Campbell}\ \emph {et~al.}(1993)\citenamefont
  {Campbell}, \citenamefont {Kaloper}, \citenamefont {Madden},\ and\
  \citenamefont {Olive}}]{Campbell:1992hc}%
  \BibitemOpen
  \bibfield  {author} {\bibinfo {author} {\bibfnamefont {B.~A.}\ \bibnamefont
  {Campbell}}, \bibinfo {author} {\bibfnamefont {N.}~\bibnamefont {Kaloper}},
  \bibinfo {author} {\bibfnamefont {R.}~\bibnamefont {Madden}}, \ and\ \bibinfo
  {author} {\bibfnamefont {K.~A.}\ \bibnamefont {Olive}},\ }\href {\doibase
  10.1016/0550-3213(93)90620-5} {\bibfield  {journal} {\bibinfo  {journal}
  {Nucl. Phys.}\ }\textbf {\bibinfo {volume} {B399}},\ \bibinfo {pages} {137}
  (\bibinfo {year} {1993})},\ \Eprint {http://arxiv.org/abs/hep-th/9301129}
  {arXiv:hep-th/9301129 [hep-th]} \BibitemShut {NoStop}%
\bibitem [{\citenamefont {Pani}\ \emph
  {et~al.}(2011{\natexlab{a}})\citenamefont {Pani}, \citenamefont {Cardoso},\
  and\ \citenamefont {Gualtieri}}]{Pani:2011xj}%
  \BibitemOpen
  \bibfield  {author} {\bibinfo {author} {\bibfnamefont {P.}~\bibnamefont
  {Pani}}, \bibinfo {author} {\bibfnamefont {V.}~\bibnamefont {Cardoso}}, \
  and\ \bibinfo {author} {\bibfnamefont {L.}~\bibnamefont {Gualtieri}},\ }\href
  {\doibase 10.1103/PhysRevD.83.104048} {\bibfield  {journal} {\bibinfo
  {journal} {Phys. Rev.}\ }\textbf {\bibinfo {volume} {D83}},\ \bibinfo {pages}
  {104048} (\bibinfo {year} {2011}{\natexlab{a}})},\ \Eprint
  {http://arxiv.org/abs/1104.1183} {arXiv:1104.1183 [gr-qc]} \BibitemShut
  {NoStop}%
\bibitem [{\citenamefont {Yunes}\ and\ \citenamefont
  {Stein}(2011)}]{Yunes:2011we}%
  \BibitemOpen
  \bibfield  {author} {\bibinfo {author} {\bibfnamefont {N.}~\bibnamefont
  {Yunes}}\ and\ \bibinfo {author} {\bibfnamefont {L.~C.}\ \bibnamefont
  {Stein}},\ }\href {\doibase 10.1103/PhysRevD.83.104002} {\bibfield  {journal}
  {\bibinfo  {journal} {Phys. Rev.}\ }\textbf {\bibinfo {volume} {D83}},\
  \bibinfo {pages} {104002} (\bibinfo {year} {2011})},\ \Eprint
  {http://arxiv.org/abs/1101.2921} {arXiv:1101.2921 [gr-qc]} \BibitemShut
  {NoStop}%
\bibitem [{\citenamefont {Pani}\ \emph
  {et~al.}(2011{\natexlab{b}})\citenamefont {Pani}, \citenamefont {Macedo},
  \citenamefont {Crispino},\ and\ \citenamefont {Cardoso}}]{Pani:2011gy}%
  \BibitemOpen
  \bibfield  {author} {\bibinfo {author} {\bibfnamefont {P.}~\bibnamefont
  {Pani}}, \bibinfo {author} {\bibfnamefont {C.~F.~B.}\ \bibnamefont {Macedo}},
  \bibinfo {author} {\bibfnamefont {L.~C.~B.}\ \bibnamefont {Crispino}}, \ and\
  \bibinfo {author} {\bibfnamefont {V.}~\bibnamefont {Cardoso}},\ }\href
  {\doibase 10.1103/PhysRevD.84.087501} {\bibfield  {journal} {\bibinfo
  {journal} {Phys. Rev.}\ }\textbf {\bibinfo {volume} {D84}},\ \bibinfo {pages}
  {087501} (\bibinfo {year} {2011}{\natexlab{b}})},\ \Eprint
  {http://arxiv.org/abs/1109.3996} {arXiv:1109.3996 [gr-qc]} \BibitemShut
  {NoStop}%
\bibitem [{\citenamefont {Zouros}\ and\ \citenamefont
  {Eardley}(1979)}]{Zouros:1979iw}%
  \BibitemOpen
  \bibfield  {author} {\bibinfo {author} {\bibfnamefont {T.~J.~M.}\
  \bibnamefont {Zouros}}\ and\ \bibinfo {author} {\bibfnamefont {D.~M.}\
  \bibnamefont {Eardley}},\ }\href {\doibase 10.1016/0003-4916(79)90237-9}
  {\bibfield  {journal} {\bibinfo  {journal} {Annals Phys.}\ }\textbf {\bibinfo
  {volume} {118}},\ \bibinfo {pages} {139} (\bibinfo {year}
  {1979})}\BibitemShut {NoStop}%
\bibitem [{\citenamefont {Dolan}(2007)}]{Dolan:2007mj}%
  \BibitemOpen
  \bibfield  {author} {\bibinfo {author} {\bibfnamefont {S.~R.}\ \bibnamefont
  {Dolan}},\ }\href {\doibase 10.1103/PhysRevD.76.084001} {\bibfield  {journal}
  {\bibinfo  {journal} {Phys. Rev.}\ }\textbf {\bibinfo {volume} {D76}},\
  \bibinfo {pages} {084001} (\bibinfo {year} {2007})},\ \Eprint
  {http://arxiv.org/abs/0705.2880} {arXiv:0705.2880 [gr-qc]} \BibitemShut
  {NoStop}%
\bibitem [{\citenamefont {Grumiller}\ and\ \citenamefont
  {Yunes}(2008)}]{Grumiller:2007rv}%
  \BibitemOpen
  \bibfield  {author} {\bibinfo {author} {\bibfnamefont {D.}~\bibnamefont
  {Grumiller}}\ and\ \bibinfo {author} {\bibfnamefont {N.}~\bibnamefont
  {Yunes}},\ }\href {\doibase 10.1103/PhysRevD.77.044015} {\bibfield  {journal}
  {\bibinfo  {journal} {Phys. Rev.}\ }\textbf {\bibinfo {volume} {D77}},\
  \bibinfo {pages} {044015} (\bibinfo {year} {2008})},\ \Eprint
  {http://arxiv.org/abs/0711.1868} {arXiv:0711.1868 [gr-qc]} \BibitemShut
  {NoStop}%
\bibitem [{\citenamefont {Regge}\ and\ \citenamefont
  {Wheeler}(1957)}]{Regge:1957td}%
  \BibitemOpen
  \bibfield  {author} {\bibinfo {author} {\bibfnamefont {T.}~\bibnamefont
  {Regge}}\ and\ \bibinfo {author} {\bibfnamefont {J.~A.}\ \bibnamefont
  {Wheeler}},\ }\href {\doibase 10.1103/PhysRev.108.1063} {\bibfield  {journal}
  {\bibinfo  {journal} {Phys. Rev.}\ }\textbf {\bibinfo {volume} {108}},\
  \bibinfo {pages} {1063} (\bibinfo {year} {1957})}\BibitemShut {NoStop}%
\bibitem [{\citenamefont {Zerilli}(1970)}]{Zerilli:1970se}%
  \BibitemOpen
  \bibfield  {author} {\bibinfo {author} {\bibfnamefont {F.~J.}\ \bibnamefont
  {Zerilli}},\ }\href {\doibase 10.1103/PhysRevLett.24.737} {\bibfield
  {journal} {\bibinfo  {journal} {Phys. Rev. Lett.}\ }\textbf {\bibinfo
  {volume} {24}},\ \bibinfo {pages} {737} (\bibinfo {year} {1970})}\BibitemShut
  {NoStop}%
\bibitem [{\citenamefont {Molina}\ \emph {et~al.}(2010)\citenamefont {Molina},
  \citenamefont {Pani}, \citenamefont {Cardoso},\ and\ \citenamefont
  {Gualtieri}}]{Molina:2010fb}%
  \BibitemOpen
  \bibfield  {author} {\bibinfo {author} {\bibfnamefont {C.}~\bibnamefont
  {Molina}}, \bibinfo {author} {\bibfnamefont {P.}~\bibnamefont {Pani}},
  \bibinfo {author} {\bibfnamefont {V.}~\bibnamefont {Cardoso}}, \ and\
  \bibinfo {author} {\bibfnamefont {L.}~\bibnamefont {Gualtieri}},\ }\href
  {\doibase 10.1103/PhysRevD.81.124021} {\bibfield  {journal} {\bibinfo
  {journal} {Phys. Rev.}\ }\textbf {\bibinfo {volume} {D81}},\ \bibinfo {pages}
  {124021} (\bibinfo {year} {2010})},\ \Eprint {http://arxiv.org/abs/1004.4007}
  {arXiv:1004.4007 [gr-qc]} \BibitemShut {NoStop}%
\bibitem [{\citenamefont {Berti}\ \emph
  {et~al.}(2009{\natexlab{a}})\citenamefont {Berti}, \citenamefont {Cardoso},\
  and\ \citenamefont {Starinets}}]{Berti:2009kk}%
  \BibitemOpen
  \bibfield  {author} {\bibinfo {author} {\bibfnamefont {E.}~\bibnamefont
  {Berti}}, \bibinfo {author} {\bibfnamefont {V.}~\bibnamefont {Cardoso}}, \
  and\ \bibinfo {author} {\bibfnamefont {A.~O.}\ \bibnamefont {Starinets}},\
  }\href {\doibase 10.1088/0264-9381/26/16/163001} {\bibfield  {journal}
  {\bibinfo  {journal} {Class. Quant. Grav.}\ }\textbf {\bibinfo {volume}
  {26}},\ \bibinfo {pages} {163001} (\bibinfo {year} {2009}{\natexlab{a}})},\
  \Eprint {http://arxiv.org/abs/0905.2975} {arXiv:0905.2975 [gr-qc]}
  \BibitemShut {NoStop}%
\bibitem [{\citenamefont {Deruelle}\ and\ \citenamefont
  {Ruffini}(1974)}]{Deruelle:1974zy}%
  \BibitemOpen
  \bibfield  {author} {\bibinfo {author} {\bibfnamefont {N.}~\bibnamefont
  {Deruelle}}\ and\ \bibinfo {author} {\bibfnamefont {R.}~\bibnamefont
  {Ruffini}},\ }\href {\doibase 10.1016/0370-2693(74)90119-1} {\bibfield
  {journal} {\bibinfo  {journal} {Phys. Lett.}\ }\textbf {\bibinfo {volume}
  {52B}},\ \bibinfo {pages} {437} (\bibinfo {year} {1974})}\BibitemShut
  {NoStop}%
\bibitem [{\citenamefont {Damour}\ \emph {et~al.}(1976)\citenamefont {Damour},
  \citenamefont {Deruelle},\ and\ \citenamefont {Ruffini}}]{Damour:1976kh}%
  \BibitemOpen
  \bibfield  {author} {\bibinfo {author} {\bibfnamefont {T.}~\bibnamefont
  {Damour}}, \bibinfo {author} {\bibfnamefont {N.}~\bibnamefont {Deruelle}}, \
  and\ \bibinfo {author} {\bibfnamefont {R.}~\bibnamefont {Ruffini}},\ }\href
  {\doibase 10.1007/BF02725534} {\bibfield  {journal} {\bibinfo  {journal}
  {Lett. Nuovo Cim.}\ }\textbf {\bibinfo {volume} {15}},\ \bibinfo {pages}
  {257} (\bibinfo {year} {1976})}\BibitemShut {NoStop}%
\bibitem [{\citenamefont {Kimura}(2018)}]{Kimura:2018nxk}%
  \BibitemOpen
  \bibfield  {author} {\bibinfo {author} {\bibfnamefont {M.}~\bibnamefont
  {Kimura}},\ }\href {\doibase 10.1103/PhysRevD.98.024048} {\bibfield
  {journal} {\bibinfo  {journal} {Phys. Rev.}\ }\textbf {\bibinfo {volume}
  {D98}},\ \bibinfo {pages} {024048} (\bibinfo {year} {2018})},\ \Eprint
  {http://arxiv.org/abs/1807.05029} {arXiv:1807.05029 [gr-qc]} \BibitemShut
  {NoStop}%
\bibitem [{\citenamefont {Cardoso}\ and\ \citenamefont
  {Gualtieri}(2009)}]{Cardoso:2009pk}%
  \BibitemOpen
  \bibfield  {author} {\bibinfo {author} {\bibfnamefont {V.}~\bibnamefont
  {Cardoso}}\ and\ \bibinfo {author} {\bibfnamefont {L.}~\bibnamefont
  {Gualtieri}},\ }\href {\doibase 10.1103/PhysRevD.81.089903,
  10.1103/PhysRevD.80.064008} {\bibfield  {journal} {\bibinfo  {journal} {Phys.
  Rev.}\ }\textbf {\bibinfo {volume} {D80}},\ \bibinfo {pages} {064008}
  (\bibinfo {year} {2009})},\ \bibinfo {note} {[Erratum: Phys.
  Rev.D81,089903(2010)]},\ \Eprint {http://arxiv.org/abs/0907.5008}
  {arXiv:0907.5008 [gr-qc]} \BibitemShut {NoStop}%
\bibitem [{\citenamefont {Rosa}\ and\ \citenamefont
  {Dolan}(2012)}]{Rosa:2011my}%
  \BibitemOpen
  \bibfield  {author} {\bibinfo {author} {\bibfnamefont {J.~G.}\ \bibnamefont
  {Rosa}}\ and\ \bibinfo {author} {\bibfnamefont {S.~R.}\ \bibnamefont
  {Dolan}},\ }\href {\doibase 10.1103/PhysRevD.85.044043} {\bibfield  {journal}
  {\bibinfo  {journal} {Phys. Rev.}\ }\textbf {\bibinfo {volume} {D85}},\
  \bibinfo {pages} {044043} (\bibinfo {year} {2012})},\ \Eprint
  {http://arxiv.org/abs/1110.4494} {arXiv:1110.4494 [hep-th]} \BibitemShut
  {NoStop}%
\bibitem [{\citenamefont {Chandrasekhar}\ and\ \citenamefont
  {Detweiler}(1975)}]{Chandrasekhar:1975zza}%
  \BibitemOpen
  \bibfield  {author} {\bibinfo {author} {\bibfnamefont {S.}~\bibnamefont
  {Chandrasekhar}}\ and\ \bibinfo {author} {\bibfnamefont {S.~L.}\ \bibnamefont
  {Detweiler}},\ }\href {\doibase 10.1098/rspa.1975.0112} {\bibfield  {journal}
  {\bibinfo  {journal} {Proc. Roy. Soc. Lond.}\ }\textbf {\bibinfo {volume}
  {A344}},\ \bibinfo {pages} {441} (\bibinfo {year} {1975})}\BibitemShut
  {NoStop}%
\bibitem [{\citenamefont {Pani}(2013)}]{Pani:2013pma}%
  \BibitemOpen
  \bibfield  {author} {\bibinfo {author} {\bibfnamefont {P.}~\bibnamefont
  {Pani}},\ }\bibfield  {booktitle} {\emph {\bibinfo {booktitle} {{Proceedings,
  Spring School on Numerical Relativity and High Energy Physics (NR/HEP2):
  Lisbon, Portugal, March 11-14, 2013}}},\ }\href {\doibase
  10.1142/S0217751X13400186} {\bibfield  {journal} {\bibinfo  {journal} {Int.
  J. Mod. Phys.}\ }\textbf {\bibinfo {volume} {A28}},\ \bibinfo {pages}
  {1340018} (\bibinfo {year} {2013})},\ \Eprint
  {http://arxiv.org/abs/1305.6759} {arXiv:1305.6759 [gr-qc]} \BibitemShut
  {NoStop}%
\bibitem [{\citenamefont {Macedo}\ \emph {et~al.}(2016)\citenamefont {Macedo},
  \citenamefont {Cardoso}, \citenamefont {Crispino},\ and\ \citenamefont
  {Pani}}]{Macedo:2016wgh}%
  \BibitemOpen
  \bibfield  {author} {\bibinfo {author} {\bibfnamefont {C.~F.~B.}\
  \bibnamefont {Macedo}}, \bibinfo {author} {\bibfnamefont {V.}~\bibnamefont
  {Cardoso}}, \bibinfo {author} {\bibfnamefont {L.~C.~B.}\ \bibnamefont
  {Crispino}}, \ and\ \bibinfo {author} {\bibfnamefont {P.}~\bibnamefont
  {Pani}},\ }\href {\doibase 10.1103/PhysRevD.93.064053} {\bibfield  {journal}
  {\bibinfo  {journal} {Phys. Rev.}\ }\textbf {\bibinfo {volume} {D93}},\
  \bibinfo {pages} {064053} (\bibinfo {year} {2016})},\ \Eprint
  {http://arxiv.org/abs/1603.02095} {arXiv:1603.02095 [gr-qc]} \BibitemShut
  {NoStop}%
\bibitem [{\citenamefont {Bl\'azquez-Salcedo}\ \emph
  {et~al.}(2016{\natexlab{a}})\citenamefont {Bl\'azquez-Salcedo}, \citenamefont
  {Macedo}, \citenamefont {Cardoso}, \citenamefont {Ferrari}, \citenamefont
  {Gualtieri}, \citenamefont {Khoo}, \citenamefont {Kunz},\ and\ \citenamefont
  {Pani}}]{Blazquez-Salcedo:2016enn}%
  \BibitemOpen
  \bibfield  {author} {\bibinfo {author} {\bibfnamefont {J.~L.}\ \bibnamefont
  {Bl\'azquez-Salcedo}}, \bibinfo {author} {\bibfnamefont {C.~F.~B.}\
  \bibnamefont {Macedo}}, \bibinfo {author} {\bibfnamefont {V.}~\bibnamefont
  {Cardoso}}, \bibinfo {author} {\bibfnamefont {V.}~\bibnamefont {Ferrari}},
  \bibinfo {author} {\bibfnamefont {L.}~\bibnamefont {Gualtieri}}, \bibinfo
  {author} {\bibfnamefont {F.~S.}\ \bibnamefont {Khoo}}, \bibinfo {author}
  {\bibfnamefont {J.}~\bibnamefont {Kunz}}, \ and\ \bibinfo {author}
  {\bibfnamefont {P.}~\bibnamefont {Pani}},\ }\href {\doibase
  10.1103/PhysRevD.94.104024} {\bibfield  {journal} {\bibinfo  {journal} {Phys.
  Rev.}\ }\textbf {\bibinfo {volume} {D94}},\ \bibinfo {pages} {104024}
  (\bibinfo {year} {2016}{\natexlab{a}})},\ \Eprint
  {http://arxiv.org/abs/1609.01286} {arXiv:1609.01286 [gr-qc]} \BibitemShut
  {NoStop}%
\bibitem [{\citenamefont {Bl\'azquez-Salcedo}\ \emph
  {et~al.}(2016{\natexlab{b}})\citenamefont {Bl\'azquez-Salcedo} \emph
  {et~al.}}]{Blazquez-Salcedo:2016yka}%
  \BibitemOpen
  \bibfield  {author} {\bibinfo {author} {\bibfnamefont {J.~L.}\ \bibnamefont
  {Bl\'azquez-Salcedo}} \emph {et~al.},\ }\bibfield  {booktitle} {\emph
  {\bibinfo {booktitle} {{Proceedings, IAU Symposium 324: New Frontiers in
  Black Hole Astrophysics: Ljubljana, Slovenia, September 12-16, 2016}}},\
  }\href {\doibase 10.1017/S1743921316012965} {\bibfield  {journal} {\bibinfo
  {journal} {IAU Symp.}\ }\textbf {\bibinfo {volume} {324}},\ \bibinfo {pages}
  {265} (\bibinfo {year} {2016}{\natexlab{b}})},\ \Eprint
  {http://arxiv.org/abs/1610.09214} {arXiv:1610.09214 [gr-qc]} \BibitemShut
  {NoStop}%
\bibitem [{\citenamefont {Leaver}(1985)}]{Leaver:1985ax}%
  \BibitemOpen
  \bibfield  {author} {\bibinfo {author} {\bibfnamefont {E.~W.}\ \bibnamefont
  {Leaver}},\ }\href {\doibase 10.1098/rspa.1985.0119} {\bibfield  {journal}
  {\bibinfo  {journal} {Proc. Roy. Soc. Lond.}\ }\textbf {\bibinfo {volume}
  {A402}},\ \bibinfo {pages} {285} (\bibinfo {year} {1985})}\BibitemShut
  {NoStop}%
\bibitem [{\citenamefont {Nollert}(1993)}]{Nollert:1993zz}%
  \BibitemOpen
  \bibfield  {author} {\bibinfo {author} {\bibfnamefont {H.-P.}\ \bibnamefont
  {Nollert}},\ }\href {\doibase 10.1103/PhysRevD.47.5253} {\bibfield  {journal}
  {\bibinfo  {journal} {Phys. Rev.}\ }\textbf {\bibinfo {volume} {D47}},\
  \bibinfo {pages} {5253} (\bibinfo {year} {1993})}\BibitemShut {NoStop}%
\bibitem [{\citenamefont {Dolan}(2013)}]{Dolan:2012yt}%
  \BibitemOpen
  \bibfield  {author} {\bibinfo {author} {\bibfnamefont {S.~R.}\ \bibnamefont
  {Dolan}},\ }\href {\doibase 10.1103/PhysRevD.87.124026} {\bibfield  {journal}
  {\bibinfo  {journal} {Phys. Rev.}\ }\textbf {\bibinfo {volume} {D87}},\
  \bibinfo {pages} {124026} (\bibinfo {year} {2013})},\ \Eprint
  {http://arxiv.org/abs/1212.1477} {arXiv:1212.1477 [gr-qc]} \BibitemShut
  {NoStop}%
\bibitem [{\citenamefont {Dudley}\ and\ \citenamefont
  {Finley}(1977)}]{Dudley:1977zz}%
  \BibitemOpen
  \bibfield  {author} {\bibinfo {author} {\bibfnamefont {A.~L.}\ \bibnamefont
  {Dudley}}\ and\ \bibinfo {author} {\bibfnamefont {J.~D.}\ \bibnamefont
  {Finley}},\ }\href {\doibase 10.1103/PhysRevLett.38.1505} {\bibfield
  {journal} {\bibinfo  {journal} {Phys. Rev. Lett.}\ }\textbf {\bibinfo
  {volume} {38}},\ \bibinfo {pages} {1505} (\bibinfo {year}
  {1977})}\BibitemShut {NoStop}%
\bibitem [{\citenamefont {{Chandrasekhar}}\ and\ \citenamefont
  {{Ferrari}}(1991)}]{1991RSPSA.434..449C}%
  \BibitemOpen
  \bibfield  {author} {\bibinfo {author} {\bibfnamefont {S.}~\bibnamefont
  {{Chandrasekhar}}}\ and\ \bibinfo {author} {\bibfnamefont {V.}~\bibnamefont
  {{Ferrari}}},\ }\href {\doibase 10.1098/rspa.1991.0104} {\bibfield  {journal}
  {\bibinfo  {journal} {Proceedings of the Royal Society of London Series A}\
  }\textbf {\bibinfo {volume} {434}},\ \bibinfo {pages} {449} (\bibinfo {year}
  {1991})}\BibitemShut {NoStop}%
\bibitem [{\citenamefont {Berti}\ \emph
  {et~al.}(2009{\natexlab{b}})\citenamefont {Berti}, \citenamefont {Cardoso},\
  and\ \citenamefont {Pani}}]{Berti:2009wx}%
  \BibitemOpen
  \bibfield  {author} {\bibinfo {author} {\bibfnamefont {E.}~\bibnamefont
  {Berti}}, \bibinfo {author} {\bibfnamefont {V.}~\bibnamefont {Cardoso}}, \
  and\ \bibinfo {author} {\bibfnamefont {P.}~\bibnamefont {Pani}},\ }\href
  {\doibase 10.1103/PhysRevD.79.101501} {\bibfield  {journal} {\bibinfo
  {journal} {Phys. Rev.}\ }\textbf {\bibinfo {volume} {D79}},\ \bibinfo {pages}
  {101501} (\bibinfo {year} {2009}{\natexlab{b}})},\ \Eprint
  {http://arxiv.org/abs/0903.5311} {arXiv:0903.5311 [gr-qc]} \BibitemShut
  {NoStop}%
\bibitem [{\citenamefont {Macedo}\ \emph {et~al.}(2018)\citenamefont {Macedo},
  \citenamefont {Stratton}, \citenamefont {Dolan},\ and\ \citenamefont
  {Crispino}}]{Macedo:2018yoi}%
  \BibitemOpen
  \bibfield  {author} {\bibinfo {author} {\bibfnamefont {C.~F.~B.}\
  \bibnamefont {Macedo}}, \bibinfo {author} {\bibfnamefont {T.}~\bibnamefont
  {Stratton}}, \bibinfo {author} {\bibfnamefont {S.}~\bibnamefont {Dolan}}, \
  and\ \bibinfo {author} {\bibfnamefont {C.~B.}\ \bibnamefont {Crispino},
  \bibfnamefont {Luís}},\ }\href@noop {} {\  (\bibinfo {year} {2018})},\
  \Eprint {http://arxiv.org/abs/1807.04762} {arXiv:1807.04762 [gr-qc]}
  \BibitemShut {NoStop}%
\bibitem [{\citenamefont {Gundlach}\ \emph {et~al.}(1994)\citenamefont
  {Gundlach}, \citenamefont {Price},\ and\ \citenamefont
  {Pullin}}]{Gundlach:1993tp}%
  \BibitemOpen
  \bibfield  {author} {\bibinfo {author} {\bibfnamefont {C.}~\bibnamefont
  {Gundlach}}, \bibinfo {author} {\bibfnamefont {R.~H.}\ \bibnamefont {Price}},
  \ and\ \bibinfo {author} {\bibfnamefont {J.}~\bibnamefont {Pullin}},\ }\href
  {\doibase 10.1103/PhysRevD.49.883} {\bibfield  {journal} {\bibinfo  {journal}
  {Phys. Rev.}\ }\textbf {\bibinfo {volume} {D49}},\ \bibinfo {pages} {883}
  (\bibinfo {year} {1994})},\ \Eprint {http://arxiv.org/abs/gr-qc/9307009}
  {arXiv:gr-qc/9307009 [gr-qc]} \BibitemShut {NoStop}%
\bibitem [{\citenamefont {Koyama}\ and\ \citenamefont
  {Tomimatsu}(2001)}]{Koyama:2001ee}%
  \BibitemOpen
  \bibfield  {author} {\bibinfo {author} {\bibfnamefont {H.}~\bibnamefont
  {Koyama}}\ and\ \bibinfo {author} {\bibfnamefont {A.}~\bibnamefont
  {Tomimatsu}},\ }\href {\doibase 10.1103/PhysRevD.64.044014} {\bibfield
  {journal} {\bibinfo  {journal} {Phys. Rev.}\ }\textbf {\bibinfo {volume}
  {D64}},\ \bibinfo {pages} {044014} (\bibinfo {year} {2001})},\ \Eprint
  {http://arxiv.org/abs/gr-qc/0103086} {arXiv:gr-qc/0103086 [gr-qc]}
  \BibitemShut {NoStop}%
\bibitem [{\citenamefont {Koyama}\ and\ \citenamefont
  {Tomimatsu}(2002)}]{Koyama:2001qw}%
  \BibitemOpen
  \bibfield  {author} {\bibinfo {author} {\bibfnamefont {H.}~\bibnamefont
  {Koyama}}\ and\ \bibinfo {author} {\bibfnamefont {A.}~\bibnamefont
  {Tomimatsu}},\ }\href {\doibase 10.1103/PhysRevD.65.084031} {\bibfield
  {journal} {\bibinfo  {journal} {Phys. Rev.}\ }\textbf {\bibinfo {volume}
  {D65}},\ \bibinfo {pages} {084031} (\bibinfo {year} {2002})},\ \Eprint
  {http://arxiv.org/abs/gr-qc/0112075} {arXiv:gr-qc/0112075 [gr-qc]}
  \BibitemShut {NoStop}%
\bibitem [{\citenamefont {Burko}\ and\ \citenamefont
  {Khanna}(2004)}]{Burko:2004jn}%
  \BibitemOpen
  \bibfield  {author} {\bibinfo {author} {\bibfnamefont {L.~M.}\ \bibnamefont
  {Burko}}\ and\ \bibinfo {author} {\bibfnamefont {G.}~\bibnamefont {Khanna}},\
  }\href {\doibase 10.1103/PhysRevD.70.044018} {\bibfield  {journal} {\bibinfo
  {journal} {Phys. Rev.}\ }\textbf {\bibinfo {volume} {D70}},\ \bibinfo {pages}
  {044018} (\bibinfo {year} {2004})},\ \Eprint
  {http://arxiv.org/abs/gr-qc/0403018} {arXiv:gr-qc/0403018 [gr-qc]}
  \BibitemShut {NoStop}%
\bibitem [{\citenamefont {Witek}\ \emph {et~al.}(2013)\citenamefont {Witek},
  \citenamefont {Cardoso}, \citenamefont {Ishibashi},\ and\ \citenamefont
  {Sperhake}}]{Witek:2012tr}%
  \BibitemOpen
  \bibfield  {author} {\bibinfo {author} {\bibfnamefont {H.}~\bibnamefont
  {Witek}}, \bibinfo {author} {\bibfnamefont {V.}~\bibnamefont {Cardoso}},
  \bibinfo {author} {\bibfnamefont {A.}~\bibnamefont {Ishibashi}}, \ and\
  \bibinfo {author} {\bibfnamefont {U.}~\bibnamefont {Sperhake}},\ }\href
  {\doibase 10.1103/PhysRevD.87.043513} {\bibfield  {journal} {\bibinfo
  {journal} {Phys. Rev.}\ }\textbf {\bibinfo {volume} {D87}},\ \bibinfo {pages}
  {043513} (\bibinfo {year} {2013})},\ \Eprint {http://arxiv.org/abs/1212.0551}
  {arXiv:1212.0551 [gr-qc]} \BibitemShut {NoStop}%
\bibitem [{\citenamefont {Degollado}\ and\ \citenamefont
  {Herdeiro}(2014)}]{Degollado:2014vsa}%
  \BibitemOpen
  \bibfield  {author} {\bibinfo {author} {\bibfnamefont {J.~C.}\ \bibnamefont
  {Degollado}}\ and\ \bibinfo {author} {\bibfnamefont {C.~A.~R.}\ \bibnamefont
  {Herdeiro}},\ }\href {\doibase 10.1103/PhysRevD.90.065019} {\bibfield
  {journal} {\bibinfo  {journal} {Phys. Rev.}\ }\textbf {\bibinfo {volume}
  {D90}},\ \bibinfo {pages} {065019} (\bibinfo {year} {2014})},\ \Eprint
  {http://arxiv.org/abs/1408.2589} {arXiv:1408.2589 [gr-qc]} \BibitemShut
  {NoStop}%
\end{thebibliography}%


%
%
%
\end{document}